\documentstyle[12pt]{article}
\begin{document}
\begin{flushright}
June 1997
\end{flushright}
\begin{center}
\large{Quark Gluon Plasma Diagnostics in a Successive
Equilibrium Scenario}\\
\vskip .1cm
\small{Pradip Roy$^a$, Jan-e Alam$^a$, Sourav Sarkar$^a$ 
Bikash Sinhar$^{a,b}$,\\
and\\
 Sibaji Raha$^c$}
\vskip .1cm
{\it a) Variable Energy Cyclotron Centre, 1/AF Bidhan Nagar, Calcutta
700 064, INDIA.}\\
{\it b) Saha Institute of Nuclear Physics, 1/AF Bidhan Nagar, Calcutta
700 064, INDIA.}\\
{\it c) Bose Institute, 93/1 A. P. C. Road, Calcutta 700 009, INDIA.}
\end{center}
\parindent=20pt

\vskip .2in
\begin{center}
{\bf Abstract}
\end{center}
The relativistic Fokker Planck equation has been used to study the 
evolution of the quark distribution in the quark 
gluon phase expected to be formed in
ultra-relativistic heavy ion collisions. The effect of thermal masses
for quarks and gluons is incorporated to take account of the
in-medium properties. We find that the kinetic
equilibrium is achieved before the system reaches the critical 
temperature of quark hadron phase transition. We find that
chemical equilibrium is not achieved during this time.
We have evaluated the 
electromagnetic probes of quark gluon plasma from the 
non-equilibrated quark gluon phase and compared them with 
those in completely equilibrated scenario. The hard QCD production 
rates for the electromagnetic ejectiles as well as the heavy quark
production rates are also calculated.

\vskip .1in
\noindent
{\bf PACS:} 25.75.+r,12.38.Mh,24.60.Ky,24.85.+p

\noindent
{\bf Keywords:} Quark-Gluon Plasma, photons, dileptons, diphotons, 
Brownian motion.

\addtolength{\baselineskip}{0.6\baselineskip}
\section*{I. Introduction}

     The theory of strong interactions - Quantum Chromodynamics (QCD)
predicts that under extreme conditions of large baryon density and/or
high temperature, the hadronic system would dissolve into their
fundamental constituents, the quarks and gluons.
It is expected that the temperature and density achievable in
nucleus-nucleus collisions at ultra-relativistic energies would be
favourable for the formation of such a phase, called the
Quark-Gluon-Plasma (QGP). It is of fundamental importance to understand
whether thermodynamic equilibrium is achieved in the quark gluon system,
so as to justify the concept of ``phase''. Ideally, the transport
theory, or the kinetic theory, should provide an appropriate framework
to consider systems out of equilibrium. The application of Boltzmann
equation to relativistic quantum systems is however laced with major
difficulties\cite{elze}. The non-abelian nature of QCD has made this
problem rather formidable and as a result, the "transport equation"
approach to many body QCD has found only limited success to date.
Nonetheless, the field is in rapid progress and some important results
have already been obtained.

     The primary motivation to study the non-equilibrium
evolution of the quark-gluon system is driven from the fact that the
characteristic time scales for the partonic processes ($q, \bar q$ and $g$)
are of the order of the lifetime of the {\it putative} QGP. 
Even if the system
achieves thermodynamic equilibrium at some point of time, the study of the
pre-equilibrium aspects is important to evaluate in the sense that the 
pollutants from this era may affect the kinematical domains where one
looks for the signals of QGP. QGP diagnostics rely
quite heavily on the phase space densities and distributions of quarks
and gluons. To what extent equilibrium is achieved should obviously 
affect these signals.      

    To this end, the mechanisms governing the approach to thermalisation
in the quark-gluon system have been a very topical issue of late
\cite{geiger,wang}.
Recently a parton cascade picture which purports to study the QCD-based
space-time evolution of the partonic system has received a fair amount of
attention\cite{geiger}. The importance of the microscopic approach embodied
in such pictures notwithstanding, their applicability to actual QGP
diagnostics is still  largely unexplored or beset with numerical
difficulties. Also, the approximations inherent in the parton-cascade
model have been questioned by some authors \cite{larry}.
In this work, we thus propose to use a physically transparent,
semi-classical model to understand the evolution of the many body
quark-gluon system towards equilibrium. 

       The central theme of our approach is to exploit the well-known
result\cite{cutler} that $gg$ cross-section is considerably larger
than $qg$ or $qq$ cross-sections, primarily because of the colour factor
of gluons. It is therefore reasonable to expect that the gluons would
thermalise among themselves appreciably earlier than the whole system of
quarks , antiquarks, and gluons. The proper time $\tau_g$ at which the
gluons equilibrate is thus considerably less\cite{shuryak} than the overall
equilibration time $\tau_0$; the value of $\tau_0$ was proposed
to be of the order of 1 fm/c by Bjorken \cite{bjorken} some time ago.

     The gluons carry about 50\% of the
momentum and sea quarks only a tiny fraction. Thus, in very high energy
collisions (RHIC or LHC energies), {\it if we confine our attention 
to the central rapidity region}, it is quite natural that from $\tau_g$
onwards, the equilibrated gluons may provide a thermal heat bath for the
sea quarks (antiquarks). This picture is further justified by the fact
that the sea quark (antiquark) density is very low compared to that of
the gluons in this region\cite{shuryak2}. Thus we are left with a system where a relatively 
small mixture of non-equilibrium degrees of freedom (quarks and
antiquarks) interact with some equilibrated degree of freedom (the `gluonic'
bath); such processes are known to give rise to Brownian motion\cite{prigogine}
which is governed by the Fokker-Planck (FP) equation\cite{balescu}. 
QCD being asymptotically free, hard collisions involving large momentum
transfers are suppressed compared to soft interactions and in our
picture, thermalisation in the quark-gluon system proceeds through many
such soft collisions. The FP equation describes, semi-classically, the
evolution of the many body quark-gluon system in a kinetic theory framework.
The system under consideration is highly relativistic and presumably at
high temperatures. Therefore, account of production and annihilation of
$q \bar q$ pairs in the gluonic heat bath must be taken.

       The goal of this paper is two-fold. First, we develop the
framework of the relativistic FP equation to describe the temporal
evolution of the quark distribution function and then use this
information to estimate the effect of the pre-equilibrium era on the QGP
diagnostics. For this latter purpose, we confine our attention to the
electromagnetic probes - photons, photon pairs, and dileptons. It is well
known that these probes constitute an especially clean class of QGP
signals\cite{jane2}; by the very nature of their interaction; these
probes tend to leave the system of strongly interacting matter 
with minimal distortion of their energy/momentum. They thus carry the
information from within the reaction zone rather more effectively, not
being masked by the details of the evolution process. Obviously,
such is not the case for hadronic probes which lose the initial
information because of their strong coupling to the rest of the system.

     The structure of the paper is as follows. In Section II,
we derive the FP equation in a form which is appropriate for our present
purpose. In Section III, we discuss the production of photons,
diphotons, and dileptons from non-equilibrated scenario.
Section IV is devoted to the discussion of our results and we
conclude with a summary in Section V.

       We reiterate that our analysis is restricted to the situation
where the central rapidity region is free of baryon number (complete
transparency), a situation which is expected to be achieved in RHIC
or LHC energies. In these circumstances, the collective evolution of the
system is governed by the scaling hydrodynamics {\it {\'a} la}
Bjorken\cite{bjorken} which we tacitly assume to be the case.

\section*{II. The Fokker-Planck (FP) Equation}

   The Boltzmann equation in the relativistically covariant 
form can be written as 
\begin{equation}
p^{\mu} \partial_{\mu} f(x,p) = C \{f\} 
\end{equation}
where $p$ is the four momentum of the test quark and $f$ is its phase
space density. $C \{f\}$ is the collision term. The left hand side of 
eq. (1) is known as ``streaming term''. 
In the spirit of boost-invariance incorporated in the scaling
hydrodynamics of Bjorken, we assume that the phase space 
density of the quark is independent of $\vec x$, {\it i.e.} 
the plasma is uniform.
Under these conditions the Boltzmann equation reads 
\begin{equation}
\frac{\partial f}{\partial t} = \frac{C \{f \}}{E}=\left(\frac{\partial
f}{\partial t}\right)_{coll}
\end{equation}
We can seperate the elastic from the inelastic collision term as follows
\begin{equation}
\left(\frac{\partial f}{\partial t}\right)_{coll}=
\left(\frac{\partial f}{\partial t}\right)_{coll}^{el}
+\left(\frac{\partial f}{\partial t}\right)_{coll}^{inel}
\end{equation}
  First we consider the elastic collisions of the test quark
with other quarks, antiquarks, and gluons present in the system. For the
system under study, the gluons are in complete equilibrium, with a density
larger than the  non-thermal densities of the quarks and antiquarks 
present in the system. The rate of collisions $w(p,q)$ is given by
\begin{equation}
w(p,q) = \sum_{j=q, \bar q,g}w^j(p,q)
\end{equation}
where $w^j$ denotes the rate of collisions 
of a test quark $q$ with other participant parton $j$, {\it i.e.} 
for the reaction $jq \rightarrow jq$. In
our case the term $w^g$ of eq. (4) dominates over the other two terms,
because of the paucity of quarks and anti quarks in the system.
The expression for $w^j$ can be written as 
\begin{equation}
w^j(p,q) = \gamma_j \int \frac{d^3q}{(2\pi)^3} f_j(q) v_{rel}
\sigma^j
\end{equation}
where $\gamma_j$ is the statistical degeneracy ($2\times 8$ for gluons) 
and $f_j(q)$ is the phase space density for the species $j$; 
$v_{rel}$ is the relative velocity between the test quark $q$ and the other
participating parton $j$ and $\sigma^j$ is the relevant cross section for the
elastic collision.
 
    We assume that the transition takes place between two states having
momenta, say, $p\prime$ and $p$, where $p\prime -p$ is very small. In
terms of collision rates this means that the function $w(p,p\prime)$
is sharply peaked at $p \approx p\prime$. The right hand side of
eq. (2) can be written as 
\begin{equation}
\left(\frac{\partial f}{\partial t}\right)^{el}_{coll}
 =\int d^3k 
\left[w(p+k,k)f(p+k)-w(p,k)f(p)\right]
\end{equation}
Expanding $w(p+k,k)f(p+k)$ by Taylor series we get
\begin{equation}
w(p+k,k)f(p+k) \approx  w(p,k)f(p)+{\vec k}\cdot \frac{\partial}{\partial
 {\vec p}}(wf)
+\frac{1}{2} k_ik_j \frac{\partial^2}{\partial p_i \partial p_j}(wf)
\end{equation}
Substituting eq. (7) in eq. (6) we get,
\begin{equation}
\left(\frac{\partial f}{\partial t}\right)_{coll}^{el}
 = \int d^3k \left[{\vec k}\cdot
\frac{\partial}{\partial {\vec p}}+\frac{1}{2} k_ik_j
\frac{\partial^2}{\partial p_i \partial p_j}\right](wf)
\end{equation}
For a one dimensional problem, we can write the above equation as
\begin{equation}
\left(\frac{\partial f}{\partial t}\right)_{coll}^{el} =
\frac{\partial}{\partial p} \left[\mu_1(p) f\right]+
\frac{\partial^2}{\partial p^2} \left[\mu_2(p) f\right]
\end{equation}
where 
\begin{equation}
\mu_1(p)=\int d^3k w(p,k) k = \frac{\langle{\delta p}\rangle}
{\delta t} = \langle F \rangle
\end{equation}
$\langle F\rangle$ is the average force acting on the test particle,
and
\begin{equation}
\mu_2(p)=\frac{1}{2} \int d^3k w(p,k) k^2= 
\frac{\langle{(\delta p)^2}\rangle}{\delta t}
\end{equation}
Combining eq. (2) and eq. (9) we get 
\begin{equation}
\frac{\partial f}{\partial t} = \frac{\partial}{\partial p}
\left[\mu_1(p) f\right] + \frac{\partial^2}{\partial p^2}
\left[\mu_2(p) f\right]
\end{equation}
This is the celebrated Landau kinetic equation \cite{balescu}, 
a nonlinear
integro~-~differential equation. The nonlinearity is caused due to the
appearence of $f$ in $\mu_1(p)$ and $\mu_2(p)$ through $w(p,k)$.
It arises from the simple fact that we are studying
a collision process which involves two particles - it should,
therefore, depend on the states of the two participating particles in
the collision process and hence on the product of the two
distribution functions. As is evident from the derivation, the equation
is valid for a weakly coupled system, where the average kinetic
energy is large compared to the two particle interaction energy.

    We can attain considerable simplicity if we replace the distribution
functions of the collision partners of the test particle by their time
independent equilibrium Fermi-Dirac or Bose-Einstein distributions
(depending on the statistical nature)
in eqs. (10) and (11). Then eq. (12) reduces to a linear
partial differential equation - usually referred to as the Fokker-Planck 
equation\cite{balescu}.

To relate $\mu_1$ and $\mu_2$ with dynamical parameters relevant for 
the system under study, let us consider the classical equation of motion of a 
particle executing Brownian motion in a heat bath,
\begin{equation}
\frac{dp}{dt} = F(t)
\end{equation}
where $F(t) = F_r(t) + F_d(t)$. Here $F_r(t)$ is the rapidly
fluctuating part and $F_d(t)$ is the viscous force. Taking the average of
the above equation and assuming that the average of 
$F_r(t)$ over a long interval of time vanishes, {\it i.e.} 
\begin{eqnarray}
\langle {F_r(t)}\rangle&=&0\nonumber\\
\langle {F_d(t)} \rangle&=&-a_pv
\end{eqnarray}
where $a_p$ is the dynamical friction parameter containing the
dynamics of the problem under study (see next section for details),
we can, in the relativistic case 
($v = p/\sqrt{p^2+m^2}$), write
\begin{equation}
\mu_1(p) = -a_pv = \frac{-a_p p}{\sqrt{p^2+m^2}}
\end{equation}
Assuming that the coupling between the Brownian particle and the bath
is weak, we have
\begin{equation}
\mu_2(p) = \frac{\langle {(\delta p)^2} \rangle}{\delta t}
=2a_p(v p)
\end{equation}
For the ultra-relativistic case $v \sim 1$, $p \sim T$, implying
\begin{equation}
\mu_2(p) {\simeq 2a_pT} {\equiv} 2D_F
\end{equation}
Using eqs. (12), (15) and (17) we get 
\begin{equation}
\frac{\partial f}{\partial t}=\frac{\partial}{\partial p}
\left(\frac{a_p p f}{\sqrt{p^2+m^2}} \right)+D_F\frac{\partial^2
f}{\partial p^2}
\end{equation}
This is the Fokker Planck equation describing the evolution 
of a quark towards equilibrium due to its interaction with 
the gluonic heat bath (see ref. \cite{somenath,hwa}.)

The relativistic FP equation with inelastic collisions 
can be written as 
\begin{equation}
\frac{\partial f}{\partial t}-\frac{\partial}{\partial {p_z}}
\left(\frac{a_p p_z f}{\sqrt{p_z^2+m_T^2}} \right)-D_F\frac{\partial^2
f}{\partial p_z^2} = \left(\frac{\partial f}{\partial t}
\right)_{coll}^{inel}
\end{equation}
We can linearize the above equation  
with the relaxation time approximation \cite{baym,heiselberg,wong} as follows,
\begin{equation}
\left(\frac{\partial f}{\partial t}\right)_{coll}^{inel} =
- \frac{f-f_{eq}}{\tau_{relax}}
\end{equation}
where $f_{eq}$ is the equilibrium distribution and $\tau_{relax}$ 
is the relaxation time estimated from the reactions
$gg\leftrightarrow q\bar q$ and $g\leftrightarrow q\bar q$,
$m_T$ is the transverse mass
($=\sqrt{p_T^2+m_{eff}^2}$). $m_{eff}$ is the effective mass defined as
\begin{equation}
m_{eff}=\sqrt{m_{current}^2+m_{thermal}^2}
\end{equation}
where $m_{current}$ is the current quark mass ($=10 $MeV for $u$ and $d$
quarks) and  $m_{thermal}$ is the thermal mass: 
\begin{equation}
m_{thermal} = \sqrt{g_s^2T^2/6} 
\end{equation}
$g_s$ is the strong coupling constant. In a chemically non-equilibrated
scenario, the thermal mass is replaced by $m_{thermal}^2=(1+r_q/2)g_s^2T^2/9$
\cite{traxler}, where $r_q$ is the ratio of equilibrium to non-equilibrium
density. We have seen \cite{jane3} that the effect of such a change on 
thermal mass has negligible effects on the final results.

The FP equation reduces to,
\begin{equation}
\frac{\partial f}{\partial t}-\frac{\partial}{\partial {p_z}}
\left(\frac{a_p p_z f}{\sqrt{p_z^2+m_T^2}} \right)-D_F\frac{\partial^2
f}{\partial p_z^2} = 
- \frac{f-f_{eq}}{\tau_{relax}}
\end{equation}

    It should be mentioned at this point although several 
authors~\cite{baym,heiselberg,wong} have used the relaxation 
time approximation to study the approach to equilibrium in a 
quark-gluon system, such a treatment is meaningful only for 
small deviations from the equilibrium configuration. We
have dwelt on the relaxation time approach in some detail 
only to clarify
the physical picture. A more consistent way is to evaluate the contribution
of the inelastic term through a time-dependent normalization of $f$, which
can be estimated by solving the momentum-integrated Boltzmann equation, 
taking proper account of the reactions $g\,\leftrightarrow\,q\,{\bar q}$ and
$g\,g\,\leftrightarrow\,q\,{\bar q}$. These details have been reported
in the literature~\cite{jane3}; for the sake of brevity, we do not repeat
them here but refer the reader to this work. It should be noted that the 
reactions $g\,g\,\leftrightarrow\,g\,g\,g...$ etc. do not appear explicitly 
as the gluons have been assumed to be thermalized so that their density is 
determined from the temperature of the bath. Also, the thermal mass of 
the gluons is an essential ingredient; otherwise the reaction 
$g\,\leftrightarrow\,q\,{\bar q}$ would be forbidden. 

\subsection*{IIa. Determination of {\bf $a_p$}}

We assume that the phase space distribution function can be
factorised as $f(\vec p,\tau)=f(p_z,\tau)G(p_T)$, where $\tau$ is the
proper time, $G(p_T)=\exp(-p_T^2/\mu^2)/\pi\mu^2$ and $f(p_z,\tau)$
is the solution of FP equation.

The friction parameter 
$a_p$ is a very crucial 
parameter. It contains the dynamics of elastic collisions ($qq, 
q \bar q $  and $qg$). It can be related to the energy loss of a 
quark in a dense partonic medium \cite{jane} in 
the following way:
\begin{equation}
a_p(p_z,\tau) = \frac{E}{p_z} \left(-\frac{dE}{dx}\right) 
\Rightarrow a_p(\tau)=\langle{\frac{E}{p_z} \left(-\frac{dE}{dx}\right)}\rangle 
\end{equation}
The approach to equilibrium for the different quark species is then determined 
by eq.(23). In principle,
$a_p$ may be determined from kinetic theory formulation of QCD through the
application of the fluctuation dissipation theorem\cite{balescu}, 
but that is indeed
far too complex a problem to handle, given the present state of the
art.
It can, however, be assumed that since the friction constant
is expected to be largely determined by the properties of the
"bath" and not so much by the nature of the test particle, one may
take $a_p(p_z,\tau) \simeq a_p(\tau)$. In this respect, we recall the earlier
work of Svetitsky\cite{svetitsky}  where the classical diffusion and drag
coefficients of a nonrelativistic charm quark propagating in a quark
gluon plasma were calculated. Although his scenario is somewhat different
from ours,
the operating equation in both cases happens to be the Fokker Planck.
In his dynamical calculations, he found approximate momentum independence
of the drag coefficient(fig.2 of \cite{svetitsky}), entirely in line with our
assumption. It is however not realistic to the values of
$a_p$ from \cite{svetitsky} for lighter quarks. We may also remark here that a recent
work has appeared in the literature \cite{selikhov} where a Fokker Planck
type equation, including the non-abelian features of QCD in the collision
terms of the transport equation, has been discussed. The main attraction
of this work is in studying the damping
of the collective colour modes, of relevance to jet quenching studies
but outside the scope of the present work.
There is also a component which governs diffusion in
momentum space, but the deviation from the abelian case is
rather small. The correction is proportional to the small
non-equilibrium deviations and as such can be
generally neglected \cite{selikhov}. 
It is however noteworthy that these authors
also relate the momentum diffusion (or friction ) constant to the partonic
$dE/dx$, as in the present work. Let us also mention that we
have assumed the temperature $T(\tau)$ to arise from the thermal bath
whereas these authors look at non-equilibrium contributions to both
$f_{g}$ and $f_{q}$ ( $f_{\overline{q}}$ ).
There have nonetheless
been some recent developments \cite{gyulassy,braaten} in connection with jet
quenching studies in QGP which may shed light on this issue.
The expression for energy loss has been calculated by various authors
in recent times \cite{gyulassy} and for light quarks, it is given by

\begin{equation}
-\frac{dE}{dx}=\frac{4\pi}{3} C_F \alpha_s^2 T^2 \ln\left(\frac{k_{max}}
{k_D}\right)\frac{1}{v^2}
\left(v+\frac{v^2-1}{2}+\ln\frac{1+v}{1-v}\right)
\end{equation}
where $\alpha_s$ is the strong coupling constant, $C_F$ is the 
Casimir invariant, $k_{max}$ is the maximal momentum ($\sim p$, the
momentum of the light quark) and $k_D$ is the Debye momentum. The
value of $\alpha_s$ is calculated from the following parameterisation
\cite{karsch}:
$\alpha_s=6\pi/(33-2n_f)ln(\kappa T/T_c)$
For heavy quarks with $E<<M_Q^2/T$  the expression for $dE/dx$ \cite{braaten}
can be written as
\begin{eqnarray}
-\frac{dE}{dx}&=&\frac{8\pi \alpha_s^2 T^2}{3}\left(1+\frac{n_f}{6}\right)
\left(\frac{1}{v}-\frac{1-v^2}{2v^2}
\ln\frac{1+v}{1-v}\right)\nonumber\\
& &\times \ln\left(2^{n_f/(6+n_f)}B(v) \frac{ET}{m_gM_Q}\right) 
\end{eqnarray}
For $E>>M_Q^2/T$ we have
\begin{equation}
-\frac{dE}{dx}=\frac{8\pi \alpha_s^2 T^2}{3}\left(1+\frac{n_f}{6}\right)
 \ln\left(2^{n_f/2(6+n_f)}0.920 \frac{(ET)^{1/2}}{m_g}\right) 
\end{equation}
where $M_Q$ is the mass of the heavy quark, $B(v)$ is a smooth 
function of velocity (see Ref.\cite{braaten} for details), $m_g$
is the thermal gluon mass, $m_g=g_s^2T^2/3(1+N_f/6)$. In the region
of $E\sim M_Q^2/T$, we have used the eqs. (26) and (27) appropriately,
as in Ref.\cite{braaten}.
It is important to mention here that we have included the radiative 
energy loss  \cite{wang2} in $dE/dx$ ($dE/dx\mid_{rad.}=2\pi\alpha_s^2C_2T^2$
(modulo log terms)); however, the effect of such 
processes is rather small.  Recently, Baier {\it et al}  
\cite{baier} have calculated the energy loss due to radiative process
$(\sim\alpha_s\sqrt{Eq_D^2/l_g}\,\ln(E/q_D^2l_g))$
including the rescattering of the radiated gluons which is
ignored in \cite{wang2}, $l_g$ is the mean free path of the gluons.  
Weldon \cite{weldon} has shown that
the energy loss is proportional to $\alpha_s^2$
for radiative process as well as for elastic collisions
\cite{gyulassy} if one assumes 
 $l_g\sim\alpha_sT$.

\subsection*{IIb. Cooling of the Gluonic Heat Bath}

The gluonic heat bath is cooling due to expansion and the rate of cooling is
determined by the relativistic hydrodynamics. 
The bulk properties of the system, {\it e.g.} the cooling law etc.,
are governed
by the equilibrated degrees of freedom. In our case the cooling law is
given by the Bjorken's scaling law with appropriate modification due
to quark production through thermal gluon fusion $gg\rightarrow q\bar q$ 
and thermal gluon  decay $g\rightarrow q\bar q$. We have obtained the 
cooling law\cite{jane3} by solving the hydrodynamic equation 
 which is parameterised as $T=\alpha/\tau^{\beta}$ where $\alpha=0.4077$,
$\beta=0.355$ at LHC and $\alpha=0.33$, $\beta=0.352$ at RHIC
energies respectively. The cooling rate in Bjorken model ($\beta=1/3$) 
is slower compared to the present case where the production of quarks 
cost some energy. We would like to mention here that the effects of
transverse expansion on cooling law would be negligibly small for
the intial parameters under consideration for RHIC and LHC energies
(see ref.\cite{jane4} for details). 

\subsection*{IIc. Solution of the Fokker-Planck Equation}

     We solve the FP equation with the following initial and boundary
conditions for a quark species $j$
\begin{equation}
f_j(p_z,\tau) \stackrel{\tau \rightarrow \tau_g}{\longrightarrow}
\Delta_j \delta(p_z) 
\end{equation}
and
\begin{equation}
f_j(p_z,\tau) \stackrel{|p_z| \rightarrow \infty}{\longrightarrow} 0 
\end{equation}
The parameter $\Delta_j$ is determined from the initial density of the partons.
$\delta(p_z)$ is a rather good approximation of the low $x$ structure
function\cite{gluck}.
We should also mention here that the final outcome of the model is
insensitive to the functional form of the initial distribution
function - a typical characteristic of the Markovian process.

     We can either solve the inhomogeneous FP equation (eq.~(23)) with
the initial and boundary conditions given by eqs.~(28) and (29), 
or equivalently, solve the homogeneous FP equation with a time-dependent 
normalization for $f$ which accounts for the evolution of quark density 
as a result of the inelastic reactions $g\,\leftrightarrow\,q\,{\bar q}$ and 
$g\,g\,\,\leftrightarrow q\,{\bar q}$~\cite{jane3}. Our actual calculations
show that there is not much difference between the two approaches, insofar
as the electromagnetic signals are considered. We therefore show here 
the results for the relaxation time approach, as most other authors
in this field have worked in this framework.

     It is important to mention here that the final outcome depends on the
value of $\Delta_i$. There is a lack of consensus about the initial value
of the quark density. We take the initial values of 
quark densities from HIJING\cite{biro,hijing}. The phase-space
density of quark is larger in case of parton cascade model \cite{geiger} 
and also
in the work of Shuryak\cite{shuryak2}. In this sense our work
corresponds to a conservative situation. 
The data from RHIC and LHC should make a distinction
among various models.

\section*{III. Electromagnetic Probes}

\subsection*{IIIa. Single Photons}

    As mentioned in the introduction, photons and dileptons are the most
efficient signals for QGP. However, apart from the photons from QGP,
there are other sources of photons\cite{jane2}, {\it e.g.}, photons from hadronic
reactions, initial hard collisions of partons and hadronic decays. In
this section we shall evaluate the photons from a non-equilibrated quark
gluon system, thermalised QGP, and from initial hard collisions.

    The dominant reaction channel for single photon emission are the 
annihilation $(q \bar q \rightarrow g \gamma)$  and Compton $(q(\bar q)g
\rightarrow q(\bar q) \gamma)$. The transverse momentum distribution of
photons produced in a reaction ($1+2 \rightarrow 3+\gamma $)
is given by
\begin{eqnarray}
\frac{dN}{d^2p_T dy}&=& \frac{\cal N}{16(2\pi)^8} \pi R_A^2 \int 
\tau d\tau d\eta p_{1T} dp_{1T} d\theta_1 dp_{2T} dy_1 dy_2\nonumber\\
& &\times\,\frac{|{\cal M}|^2}{|p_{1T}\sin(\theta_1-\theta_2)
+p_T \sin \theta_2|_{\theta_2^0}}\nonumber\\
& &\times f(p_{1z},\tau)G(p_{1T})f(p_{2z},\tau)G(P_{2T})
\left(1\pm f(p_{3z},\tau)G(p_{3T})\right)
\end{eqnarray}
with
\begin{eqnarray}
\theta_2^0&=&\psi-\cos^{-1} \left(\frac{H}{2R p_{2T}}\right)\nonumber\\
R&=&(p_{1T}^2+p_T^2-2p_Tp_{1T} \cos\theta_1)^{1/2}\nonumber\\
\psi&=&\tan^{-1} \left( \frac{p_{1T} \sin \theta_1}{p_{1T}
\cos \theta_1 -p_T}\right)\nonumber\\
\end{eqnarray}
where
\begin{eqnarray}
H&=&m_1^2+m_2^2-m_3^2+2p_{1T}p_T\cos\theta_1
+2m_{1T}M_{2T}\cosh(y_1-y_2)\nonumber\\
& &-2m_{1T}p_T\cosh(y_1-y)-2m_{2T}p_T\cosh(y_2-y)\nonumber\\
\end{eqnarray}
We have written the total distribution function as
$f({\vec p},\tau)=G(p_T)f(p_z,\tau)$ \cite{bialas} where
$G(p_T)=\exp(-p_T^2/\mu^2)/{\pi \mu^2}$
with $\mu=0.42$ GeV\cite{geiger}. 

   For the photons from non-equilibrated and equilibrated QGP we take
the phase space distribution from the solution of FP equation 
and Fermi-Dirac distribution, respectively, for the quarks. 
Gluons are always described by Bose-Einstein distribution. 

The hard QCD photon spectra from the Compton process is evaluated
\cite{jane2} by using the
following expressions,

\begin{eqnarray}
\frac{d\sigma^C(y= 0)}{dyd^2p_T}&=&
 {\alpha\alpha _s\over 3s^2 (x_T/2)}
\int^{1}_{x_{min}}
{dx_a\over x_a-(x_T/2)}\nonumber\\
& &\times\left[F_2(x_a;A)G(x_b;B){x_b^2
+(x_T/2)^2\over x_a^2x_b^3}
+(x_a\leftrightarrow x_b;A\leftrightarrow B)\right]
\end{eqnarray}
where
\begin{equation}
x_b={x_ax_T\over 2x_a-x_T} , x_{\min }=
{x_T\over 2-x_T}
\end{equation}
$F_2(x)=x\sum e^2_q\left[q(x)+\bar{q}(x)\right]$ 
and $G(x) =x g(x)$,
where $g(x)$ and $q(x)$ ($\bar q(x)$ ) are the structure functions for 
gluons and quarks (antiquarks), respectively.

Similarly, the result for the annihilation graph is given by
\begin{eqnarray}
\frac{d\sigma^A(y= 0)}{dyd^2p_T}&=&{8\alpha
\alpha_s
\over 9s^2}\int^1_{x_{min}}{dx_a\over x_a-(x_T/2)}\nonumber\\
& &\left[Q_2(x_a;A)\bar{Q}(x_b;B){x_a^2 + x_b^2 \over x_a^3 x_b^3}
+(x_a\leftrightarrow x_b ;A\leftrightarrow B)\right]
\end{eqnarray}
where we have defined
$Q_2(x)=x\sum e^2_q q(x), \bar{Q}(x)=x\sum\bar{q}(x)$

The strong "running coupling constant" is given by,
\begin{equation}
\alpha _{s}(Q ^{2})= {12\pi\over (33-2 N_f)\ln (Q^2/
\Lambda_{QCD}^2)}
\end{equation}

where $N_{f}$ is the number of flavours
and $\Lambda_{QCD}$ is the QCD scale parameter. We have chosen 
$Q^2=p_T^2$ for evaluation of single photon spectra from 
hard QCD processes.

\subsection*{IIIb. Dileptons}

      The dilepton spectrum has been considered a promising probe for
the QGP diagnostics. In the dilepton mass window M $=$ 2--5 GeV, the
main source is  the eletromagnatic annihilation of quarks and
antiquarks. The invariant dilepton mass spectrum is given by 
\begin{eqnarray}
\frac{dN}{dM^2dy}&=&\frac{\pi R_A^2}{2(2\pi)^5} \int 
\tau d\tau d\eta
d\theta p_T dp_T q_T dq_T \nonumber\\
& &\times f(p_z^0,\tau)f({\bar p}_z^0,\tau)G(p_T)G({\bar p}_T)
\Gamma \sigma_{q \bar q \rightarrow \mu^+ \mu^-}
\end{eqnarray}
where
$\Gamma = |{\bar p}_0 {\vec p}-{\vec {\bar p}}p_0|/(p_z^0M_T)$.
The basic cross-section  for dilepton production due to $q\bar q$
annihilation is given by
\begin{eqnarray}
\sigma_{q \bar q \rightarrow \mu^+ \mu^-}&=&\frac{80\pi}{9}
\frac{\alpha^2}{M^2} \left(1-\frac{4m_{\mu}^2}{M^2}\right)^{1/2}
\left(1-\frac{4m_q^2}{M^2}\right)^{-1/2}\nonumber\\
& &\times \left [1+\frac{2m_{\mu}^2}{M^2}+\frac{2m_q^2}{M^2}+\frac{4m_q^2m_{\mu}^2}{M^4} \right]
\end{eqnarray}
We obtain $f(p_z,\tau)$ from the solution of FP equation with
appropriate boundary conditions as mentioned above. The dilepton 
production rate in complete thermal equilibrium (both kinetic and
chemical) is obtained by replacing the quark distribution function
by Fermi-Dirac distribution function. 

The Drell-Yan production has been obtained by using the expression
given below \cite{jane2}.
\begin{equation}
{d\sigma\over dM^2dy}={4\pi\alpha^2\over 3sN_c M^2}\sum_{f=1}^{N_f}
e_f^2\left[q^B_f(x)
\overline{q}^A_f(x)+q_f^A(x)
\overline{q}_f^B(x)\right]
\end{equation}

The structure functions $q_f(x)$ has been taken from ref.\cite{mrst1}.

\subsection*{IIIc. Diphotons}

     For kinematic purposes, it is naturally preferrable and useful
to work with a
pair of particles in the final state which would allow an invariant
mass identification and thus a filter. 
To this end, working with dileptons or diphotons 
is advantageous compared to single photons.
The disadvantage with diphotons is that it is a $\alpha^2$ process
whereas single photon is a $\alpha \alpha_s$ process in the lowest
order; consequently a substantial decrease with respect to single photon
rates. The possibility of treating diphoton as a signal of QGP was
considered in \cite{yosidha,datta}. The basic cross-section for diphoton
production is
\begin{eqnarray}
\sigma_{q\bar{q}}^{\gamma\gamma}(M)&=&2\pi\alpha^2\,N_c\,(2S+1)^2\,
\sum_q\frac{e_q^4}{M^2-4m_q^2}\nonumber\\
& &\times\left[\,\left[1+\frac{4m_q^2}{M^2}-\frac{8m_q^4}{M^4} \right]
\ln\left\{\frac{M^2}{2m_q^2}
\left[1+\left[1-\frac{4m_q^2}{M^2}\right]^{1/2}\right]-1\right\}
\right.\nonumber\\
&&-\left.\left[1+\frac{4m_q^2}{M^2}\right]\left[1-\frac{4m_q^2}{M^2}\right]
^{1/2}\right]
\end{eqnarray}
The invariant mass distribution for the diphoton is the same as eq. (37),
only the dilepton cross-section being replaced by diphoton cross-section.
We should however note that the temperature dependence of the two
cross-sections through the thermal mass is quite different.
For diphoton production from a thermalised QGP and also
from the initial hard QCD collisions, the calculation proceeds
along the same line as for dilepton production.
For the evaluation of dileptons from DY process and hard 
diphoton we have chosen $Q^2$ to be equal to $M^2$.

\section*{IV Results}

\subsection*{IVa. Approach to Equilibrium}

      As mentioned earlier, in the system under study the quark density
changes with proper time due to two mechanisms. The expansion
dynamics (flow) dilutes the density and on the other hand, 
the creation of quarks
in the relativistic heat bath enhances the quark density. 
The gluon density decreases due to expansion only.
We calculate
the ratio of the width of the distribution in non-equilibrium and 
equilibrium situations, {\it i.e.} ${\langle{p_z^2} \rangle}^{non-eq}/{\langle
{p_z^2} \rangle}^{eq}$. The advantage of calculating the ratio is that
the expansion effects will get cancelled to some extent, though the
cooling in the equilibrium and non-equilibrium scenario is
different as has been mentioned before. 

It is obvious that the
cooling in the non-equilibrium situation should be faster than Bjorken boost
invariant cooling law: $T \sim 1/\tau^{1/3}$. The reason behind the 
faster cooling in the present scenario is due to the production of
quarks in the gluonic heat bath through the reaction described above, 
soaking away energy and thus, cooling rapidly.

     In fig.~1 we plot these ratios as a function of the proper time
$\tau$.  At RHIC (LHC) the initial thermalisation time, $\tau_g$ for 
the gluons is $0.3$ fm/c ($0.25$ fm/c), the temperature  $T_g(\tau_g)$ is
$500$ MeV ($660$ MeV) and the initial quark density $n_{u/d}$ is $0.7$
fm$^{-3}$ ($2.8$ fm$^{-3}$).
At RHIC energy (fig.~1(a)) we observe that the ratio $D$ saturates to a
value $\sim$ 1, at a proper time $\tau \sim$ 3 fm/c, well before the
temperature of the system reaches to $T_c$ ($\sim$ 160 MeV). In 
fig.~1(b) we plot $D$ for LHC energies; the thermal equilibration 
is complete within the proper time $\sim$ 2 fm/c.
Fig.~2 ((a) and (b)) brings out the same information in greater
detail. We plot here the width of the momentum distribution of
quarks as well as gluons. The decrease of the width for gluons
corresponds to the cooling due to expansion. One can readily see the
width for quarks first increases and then starts falling just like
the gluons, indicating a clear onset of thermal (kinetic)
equilibrium. 

We evaluate the density of quarks in the 
non-equilibrium scenario by integrating the distribution
function $f(p_z,\tau)G(p_T)$ over its momentum.  
The non-equilibrium density
$n_q$ has an explicit dependence on $\tau$ and an implicit dependence
on $\tau$ through $T(\tau)$. But the equilibrium density $n_{eq}(T)$
has only an implicit dependence on $\tau$ through $T(\tau)$. The
ratio $r_q = n(\tau)/n_{eq}(T(\tau))$ thus assumes an 
{\it universal} feature, since the implicit
time dependence gets eliminated. The time dependence of the ratio $r_q$ can
then be used as a ready marker for chemical equilibrium; the time at which
the explicit time dependence of ${r_q}$ vanishes, simultaneously with 
$r_q \rightarrow$ 1, corresponds to the time
for chemical equilibration for the flavour $q$. We observe from
fig.~(3) that 
$r_q$ neither saturates nor approaches the value unity before the
temperature of the system reaches the value $T_c$ ($160$ MeV).
Therefore, we conclude that the chemical equlibrium is not achieved 
in the quark gluon system, although thermal(kinetic) equilibrium is.
To show the sensitivity of the evolution of $r_q$ on the initial
quark density we include, in fig.~(3), the result from one of our
previous calculations~\cite{jane3} obtained by taking initial
quark density from the structure functions~\cite{shuryak,gluck}. 
For the sake of completeness we also show the cooling law in
fig.~(3), where $q \bar q$ production has been taken into account.
We also see in fig.~(4) that the production of $s \bar s$ and $c
\bar c$ pairs from $q \bar q$ ($q = u/d$), $gg$ fusion and $g$
decay. As expected the results are not drastically altered from
the case with $q\bar q$ and $gg$ fusion.

\subsection*{IVb. Electromagnetic Signals}

      In the preequilibrium era, the temperature is very high but the quark
density is far below its fully equilibrated value. In the standard
scenario where one assumes the system is fully equilibrated at a proper
time $\tau_0$ ($\sim$ 1 fm/c), the temperature is small for a given
multiplicity, but the quark density is large. Within the framework
of our model, we study the effect of these competing aspects on the
photon, dilepton, and diphoton spectra. 
Clearly the above effect will be maximum
for the reactions which involve quark-antiquark annihilation {\it i.e.} on
dilepton and diphoton spectra.

In fig.~5 we plot the
single photon spectrum for LHC energies. The hard QCD photons dominates
the spectra for $p_T >$ 5 GeV, but these photons are under control
through perturbative QCD (pQCD). Photons from non-equilibrated 
and equilibrated QGP are not distinguishable upto a $p_T$ of 5 GeV. 
Low $p_T$ photons originate from the late stage of the evolution, 
when the temperature approaches the critical value $T_c = $ 160 MeV. 
For $p_T >$ 7 GeV, the photons from the non-equilibrated 
QGP system dominate, essentially because of the very high 
initial temperature though the system is far from chemical 
equilibrium.

    At RHIC,( fig~.6), the picture is different. 
Photons from equilibrated QGP
dominate the spectra upto $p_T =$ 8 GeV. For $p_T >$ 8 GeV
pre-equilibrium photons dominate because of the high initial
temperature. At RHIC energies the system remains far from chemical 
equilibrium, {\it i.e.}, the number of quarks and antiquarks is rarer
compared to their equilibrated value till the critical temperature
$T_c$. So the photons are less in number in the pre-equilibrium scenario 
compared to the equilibrium one. The hard photons dominate for $p_T$ 
above 5 GeV, as in the case of LHC energies. Once again they are
under control through pQCD. We have confined our attention entirely to
the annihilation and Compton channels for photon production in the quark
gluon sector. Even though the number of quarks (antiquarks) is low,
these channels still dominate over $gg \rightarrow g \gamma$, as seen
from fig.~(7).

     Figs.~(8) and (9) show the dilepton count rates at LHC and RHIC 
energies respectively. It is readily seen that at LHC (fig.~(8)),
preequilibrium dominates over equilibrium configurations for all
invariant masses. For RHIC (fig.~(9)), however, preequilibrium dominates
for $M < 2.5-3.0$ GeV and equilibrium for $M > 3.0$ GeV. This can be
understood in the following way. At LHC, the departure from equilibrium
is not so large while at RHIC, it is substantial (fig.~(3)).
Thus at LHC, the higher initial temperature, together with not too low
quark density, results in preequilibrium emission dominating over
the equilibrium scenario for all values of $M$. For RHIC energies,
however, the higher initial temperature is largely compensated
by the very low quark densities. In both cases, Drell-Yan (hard)
is seen to dominate over both preequilibrium and equilibrium emissions
for $M \geq 2$ GeV but once again, they can be taken care of through
pQCD.

     Figs.~(10) and (11) show the diphoton count rates at LHC and RHIC,
respectively. They behave qualitatively in the same way as dileptons. 
This is to be expected, since both dileptons and diphotons
come from the $q \bar q$ channels and as such, they only differ
in the elementary cross-sections (including the effects of thermal
masses).

\section*{V. Summary and Conclusions}

   We have analysed the approach to thermal and chemical equilibrium
in a quark gluon system within the framework of a semi-classical,
physically transparent model. A fundamental consequence of this picture
is that while thermal (kinetic) equilibrium is probable, chemical
equilibrium is not, even for LHC energies. Even the kinetic equilibrium
is achieved through a succession of time scales. The central issue 
of this work is to explore the possibility of testing the equilibrium
or pre-equilibrium scenario through the so-called signals of quark -
gluon plasma.

   We have therefore calculated the various electromagnetic probes for such a
successive equilibration picture. It is seen that both at RHIC and LHC
energies, emission from the preequilibrium phase does affect the
electromagnetic signals of the quark-gluon-plasma at the kinematic
windows thought to be appropriate for such studies. If the goal of heavy
ion studies is to look for fully equilibrated quark-gluon-plasma,
then dileptons and photons at $M \geq 2.5$ GeV appear to have a better
chance of being successful at RHIC. Curiously, even though the system
appears to approach an equilibrium configuration much more closely at
LHC than at RHIC, the dilepton or diphoton signals seem to perform better
at RHIC energies, at least for the above purpose. It should however be
noted that while dilepton and diphoton spectra behave very similarly,
a single photon spectrum has a markedly different behaviour. This
obviously is due to the fact that the single photon spectrum has
contribution also from the Compton ($qg \rightarrow q \gamma$) channel,
in addition to annihilation ($q \bar q \rightarrow q \gamma$) where as
dilepton or diphoton are sensitive only to annihilation. Thus a
correlated measurement of $\gamma$ and $\gamma \gamma$ would shed light
on the early evolution of the gluon density while simultaneous
measurement of $\mu^+ \mu^-$ and $\gamma \gamma$ would test the
validity of the mechanism visualised here to study the approach
to equilibrium.

Finally the present calculation $dN_{c \bar c}/dy$ at RHIC
energies ($T_g = 500$ MeV) turns out to be $\sim 0.2$ which compares
favourably with that of Shuryak ($\sim 0.3$)~\cite{shuryak}. Inclusion
of thermal masses suppresses the heavy quark production in the present case
compared to that of Shuryak's calculation where the thermal quark masses 
have been neglected. 

    We note that in Ref.~\cite{traxler} the authors observe 
photons from chemically non-equilibrated partonic gas is less than 
that from equilibrated plasma by a factor of $10^{-2} (10^{-1})$ at 
RHIC (LHC) energies for $1\,<\,p_T\,<3$ GeV, whereas Shuryak~\cite{shuryak} 
finds an enhancement in both photon and dilepton productions over the 
equilibrium scenario at RHIC energies. In Ref.~\cite{shuryak,traxler}
the kinetic equilibrium is assumed throughout the evolution history
unlike the present case. 
However, in the present case at LHC photons from non-equilibrated
quark-gluon system and fully equilibrated QGP are not distinguishable,
whereas at RHIC the equilibrium photon yield is order of
magnitude larger than that of non-equilibrated scenario. 
This is largely because of the fact that at LHC the initial partonic gas
has more chance to get closer to chemical equilibrium compared to RHIC. 
The difference in photon/dilepton spectra 
from all these calculations could be largely attributed to the initial 
conditions where much work needs to be done~\cite{larry}. 
The data from RHIC and LHC will make distinction among 
various initial conditions. 

    We gratefully acknowledge~ helpful discussions with Gerry Brown, 
Edward ~Shuryak and Larry~ McLerran at various ~phases of this work~.

\section*{Figure Captions}

\noindent
Figure~1. Ratio $(D)$ of the width  of the non-equilibrium to the
equilibrium momentum distribution for $u$ and $d$ quarks as a function
of proper time $\tau$ (a) at RHIC energies, (b) at LHC energies. 

\noindent
Figure~2. Width of the momentum distributions of quarks ($u/d$) and gluons
as a function of proper time $\tau$ (a) at RHIC energies, (b) at LHC
energies.

\noindent
Figure~3. The evolution of non-equilibrium density normalised to equlibrium
density, and temperature as a function of proper time $\tau$ 
(a) at RHIC energies, (b) at LHC energies.  

\noindent
Figure~4. Heavy quark production at RHIC and LHC energies as a function
of temperature at $y = 0$.

\noindent
Figure~5. Transverse momentum distribution of single photons at LHC
energies. 
The initial parameters for the equilibrium case are taken as
$T_i = 290$ MeV and $\tau_i = 1$ fm/c.

\noindent
Figure~6. Same as fig.~5 at RHIC energies.
The initial parameters for the equilibrium case are taken as
$T_i = 250$ MeV and $\tau_i = 1$ fm/c.

\noindent
Figure~7. Transverse momentum distribution of single photons from the
channel $gg \rightarrow g\gamma$ compared to $q \bar q \rightarrow g
\gamma$ at RHIC and LHC energies.

\noindent
Figure~8. Invariant mass distribution of muon pairs at LHC energies
with initial parameters same as fig.~(5).

\noindent
Figure~9. Same as fig.~8 for RHIC energies 
with initial parameters same as fig.~(6).

\noindent
Figure~10. Invariant mass distribution of photon pairs at LHC energies
with initial parameters same as fig.~(5).

\noindent
Figure~11. Same as fig.~10 for RHIC energies 
with initial parameters same as fig.~(6).

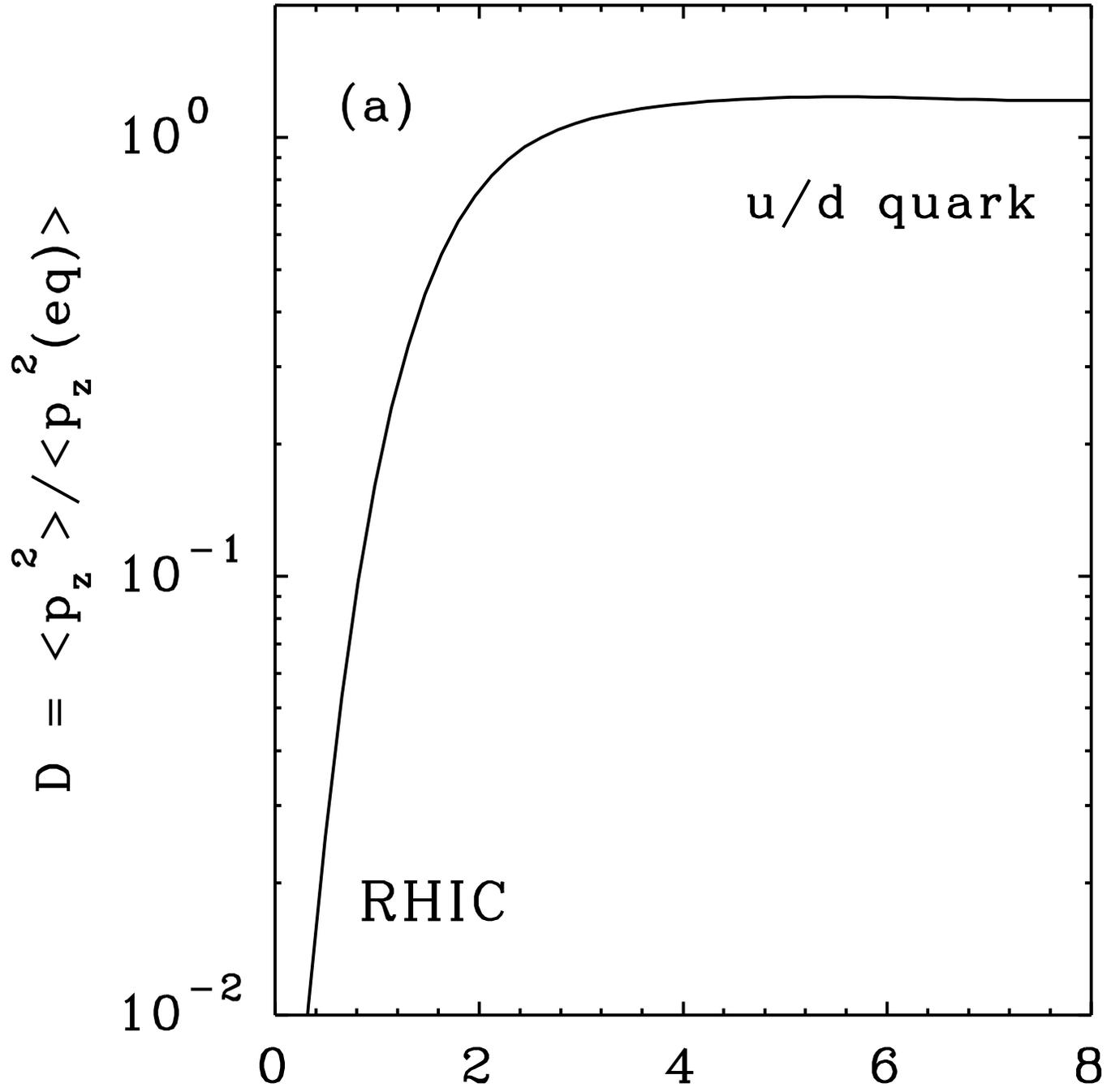
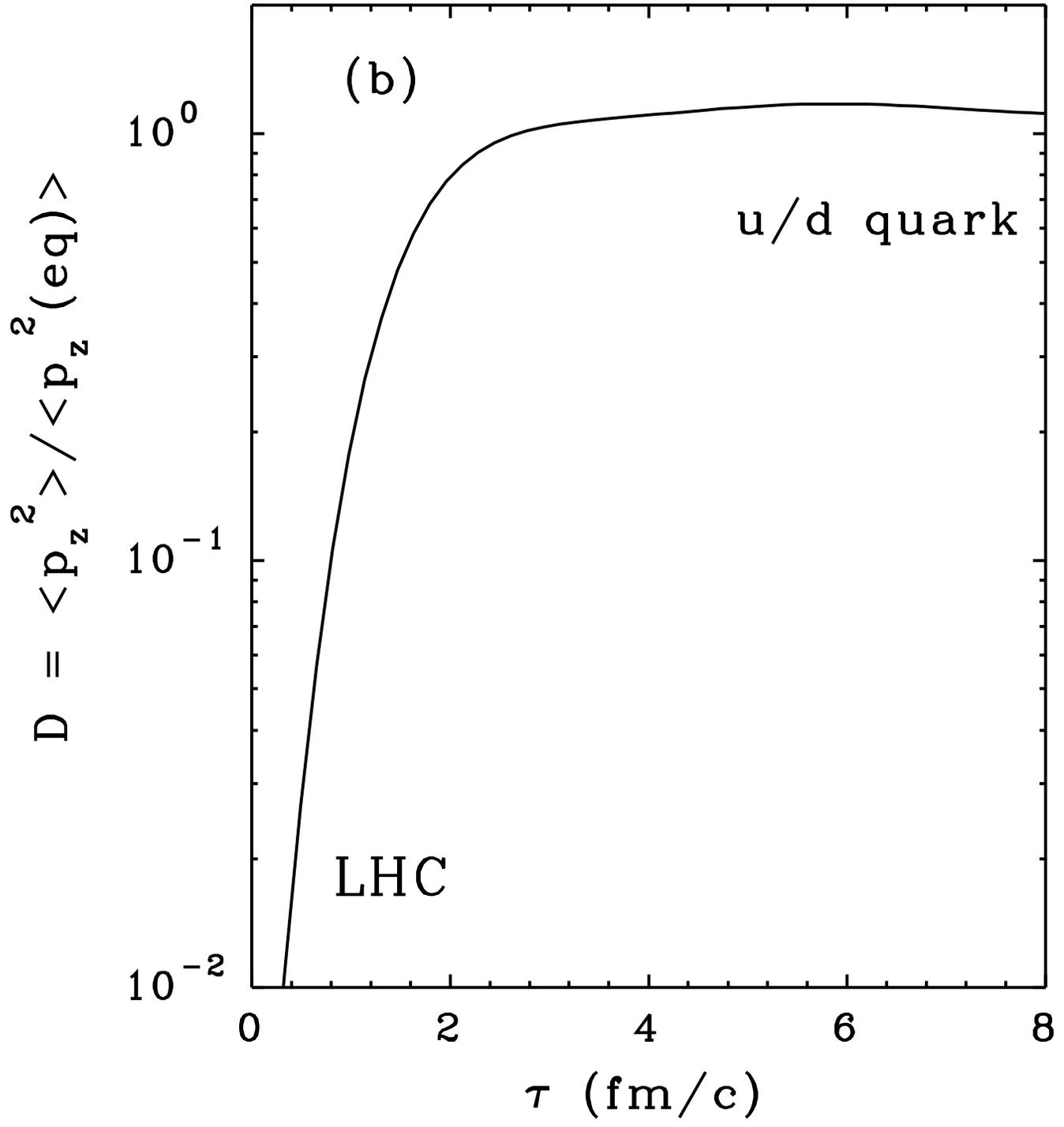
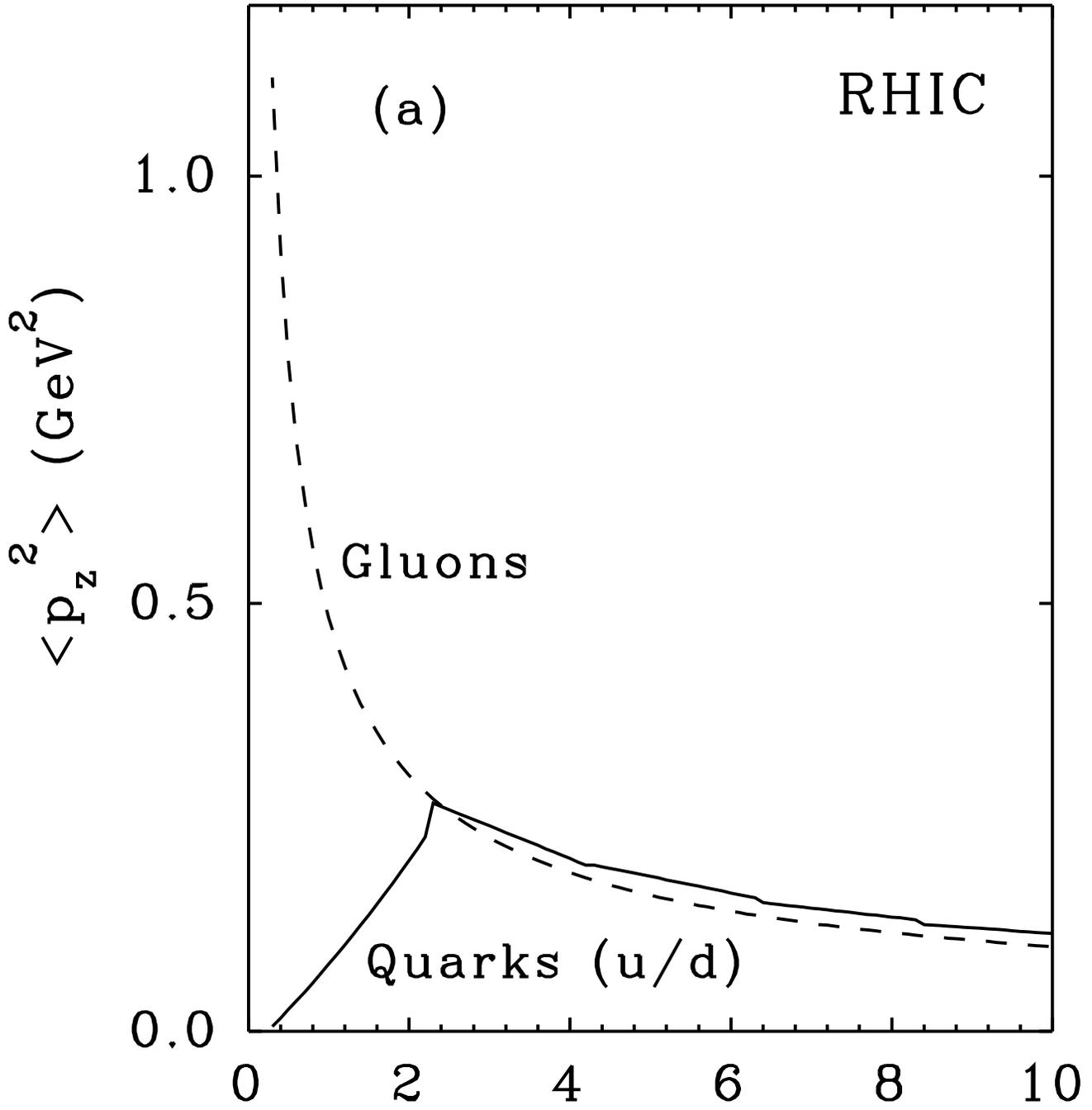
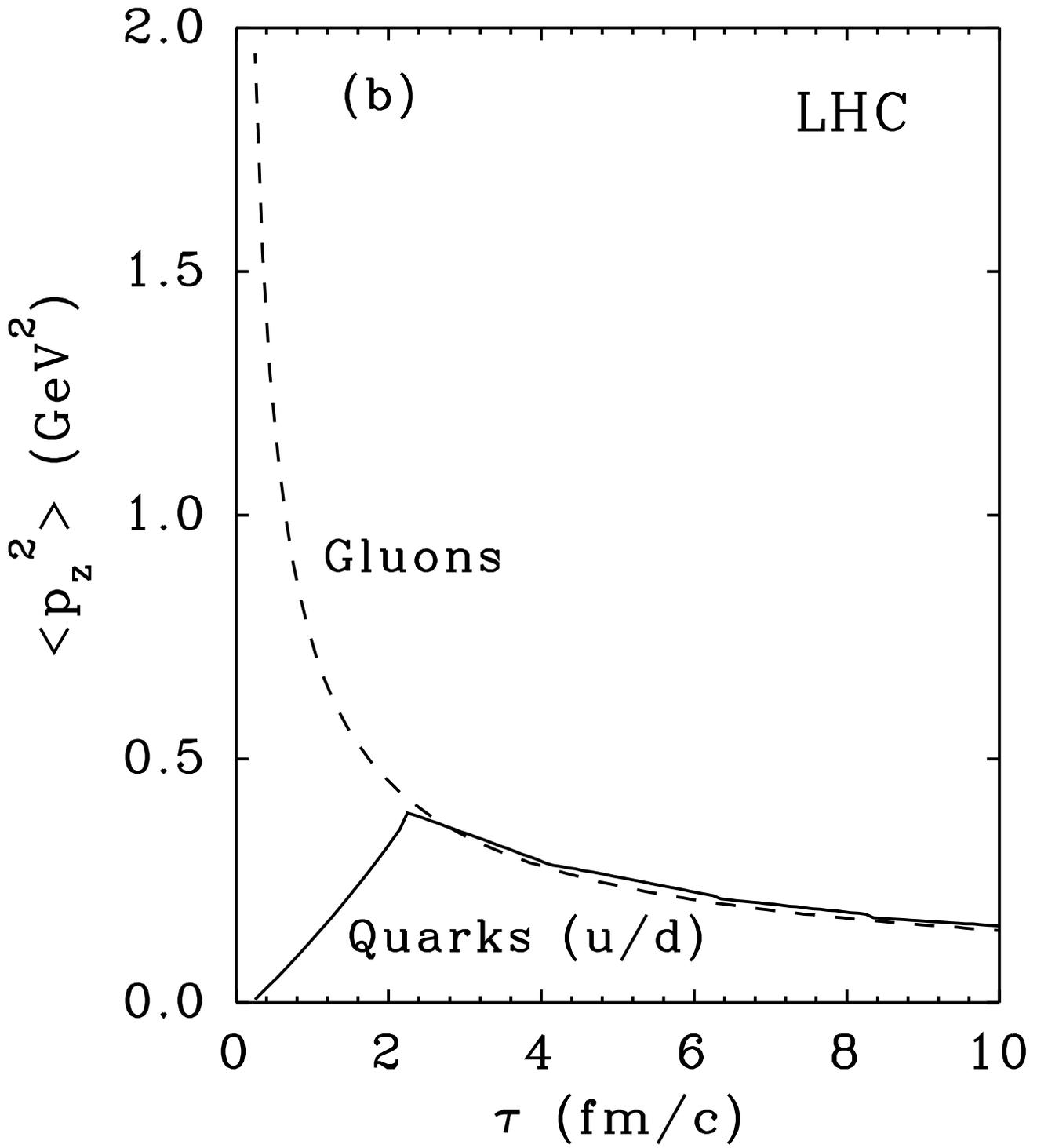
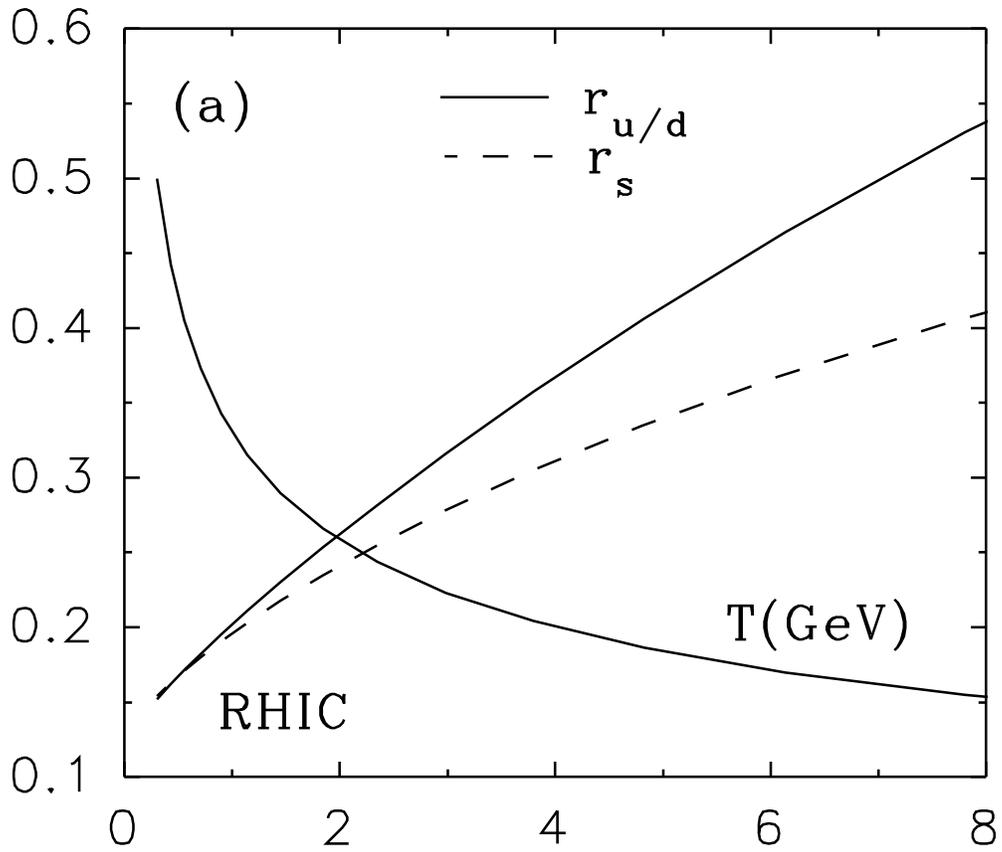
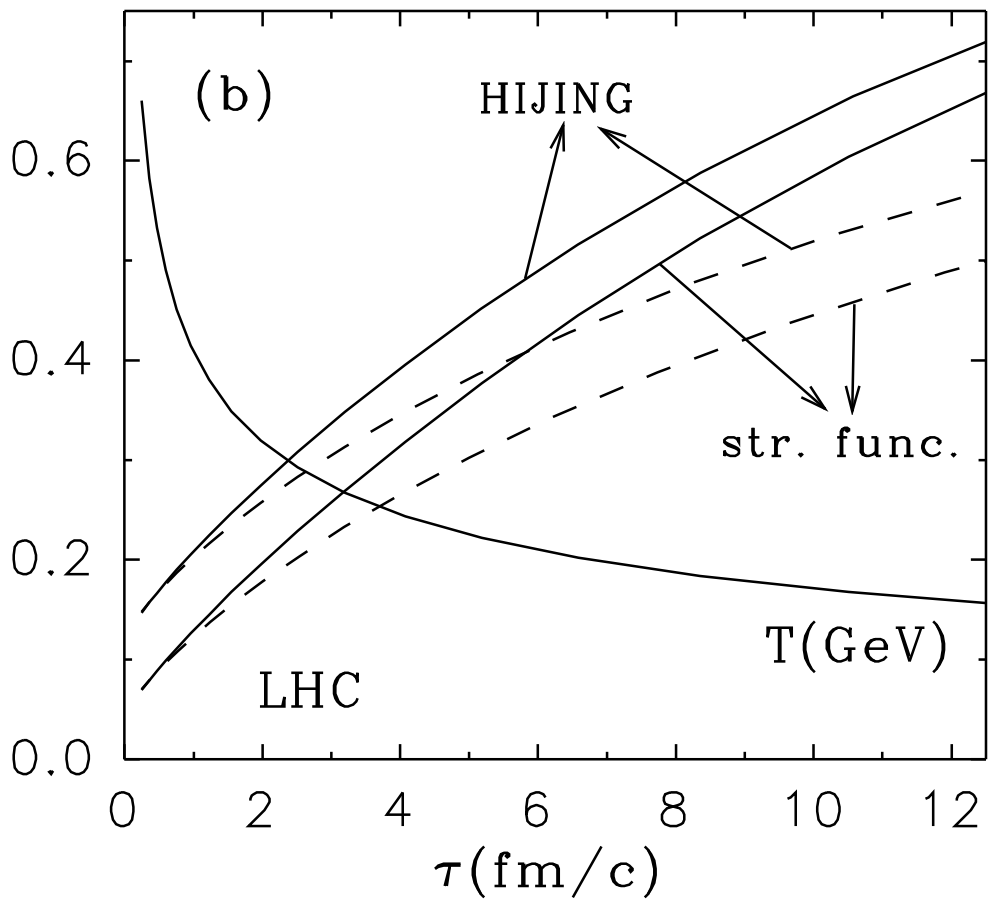
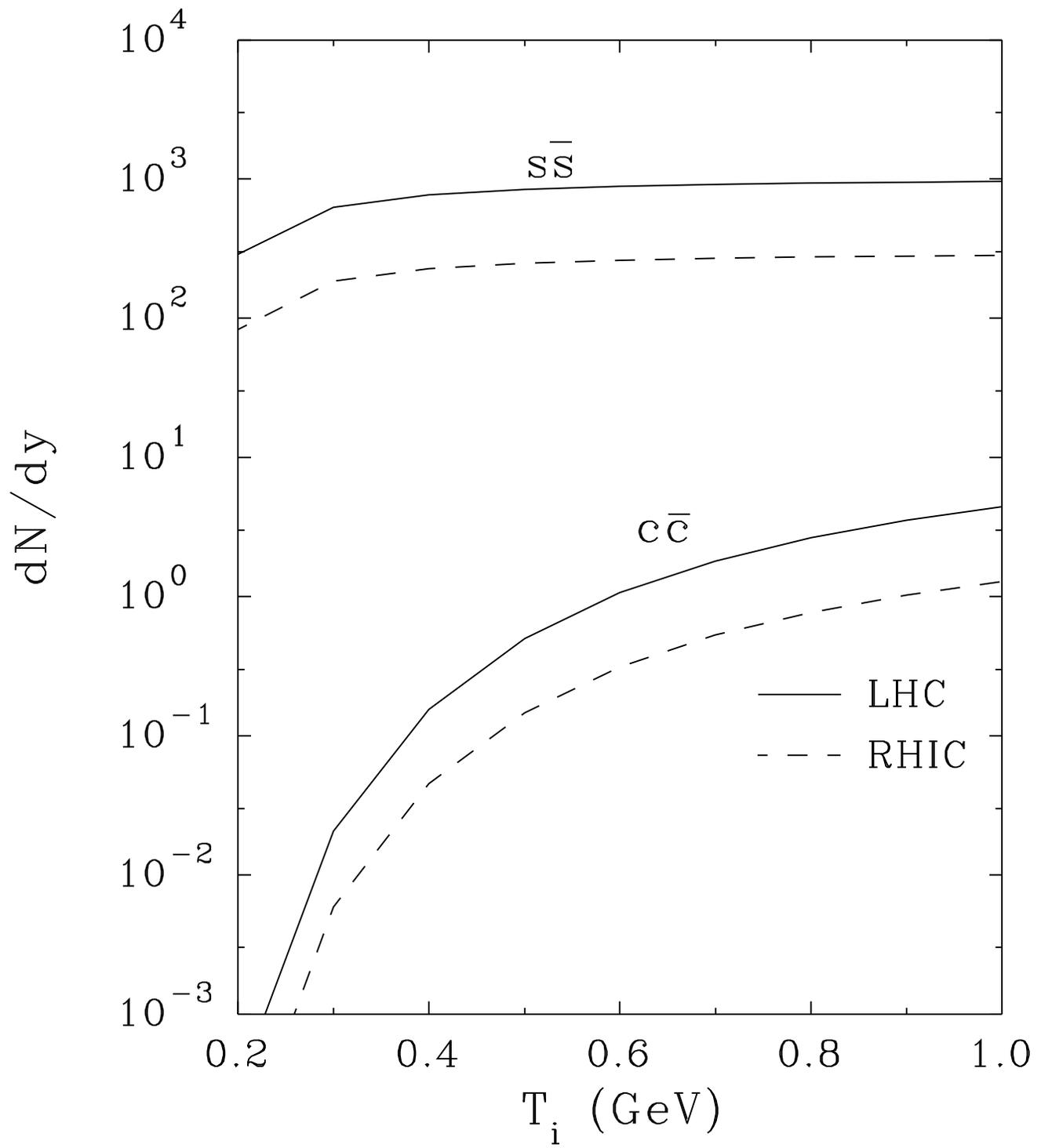
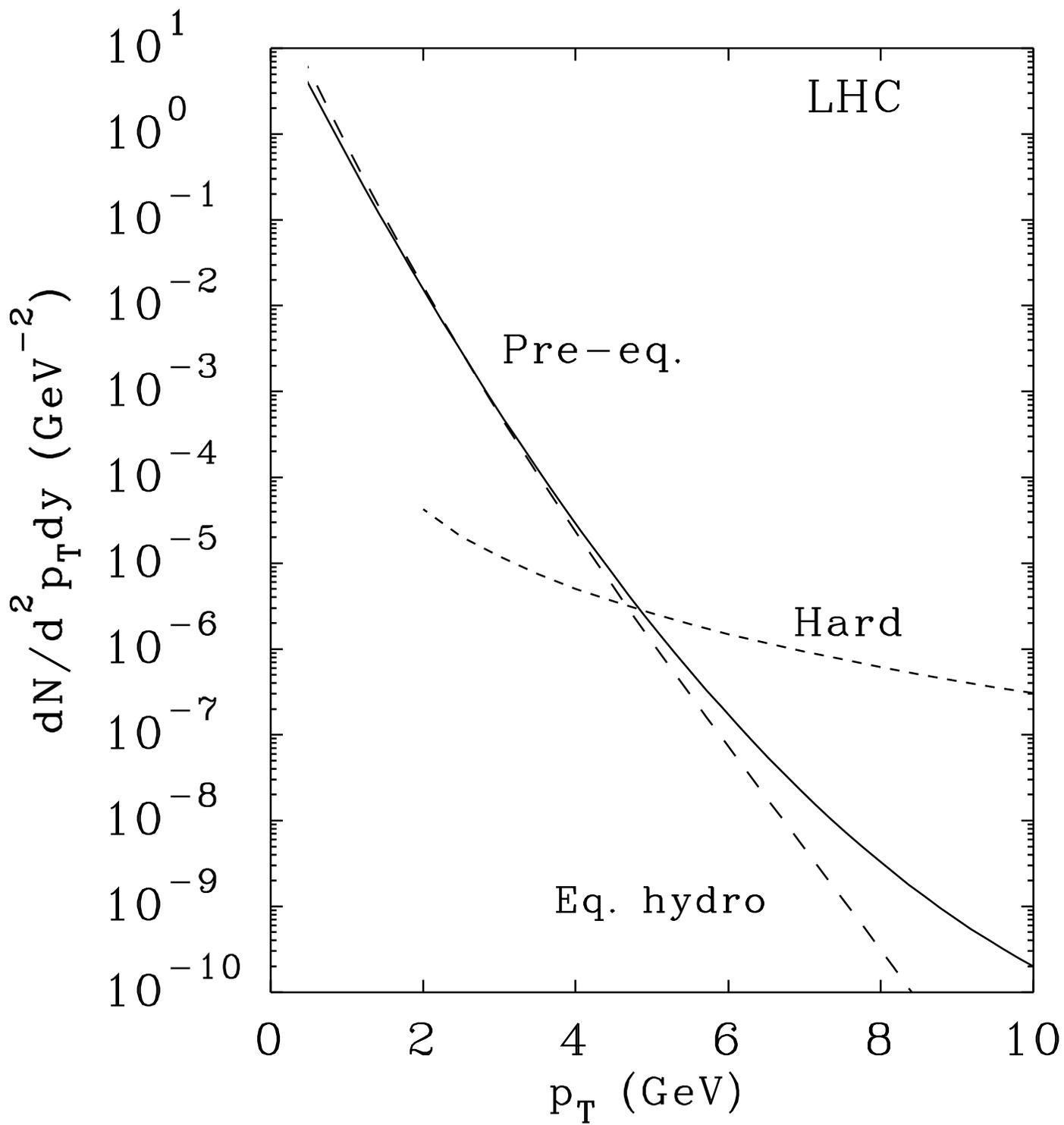
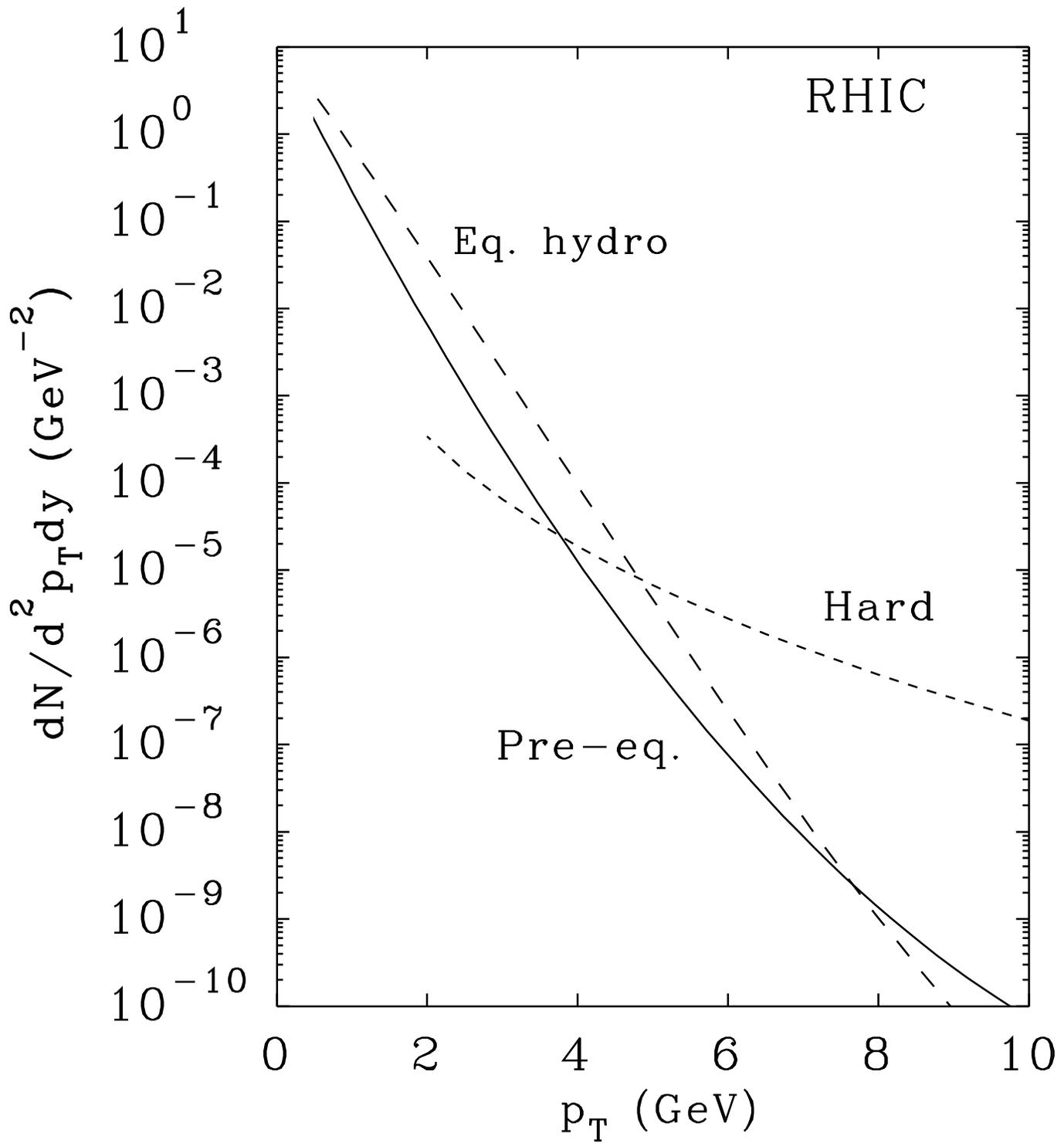
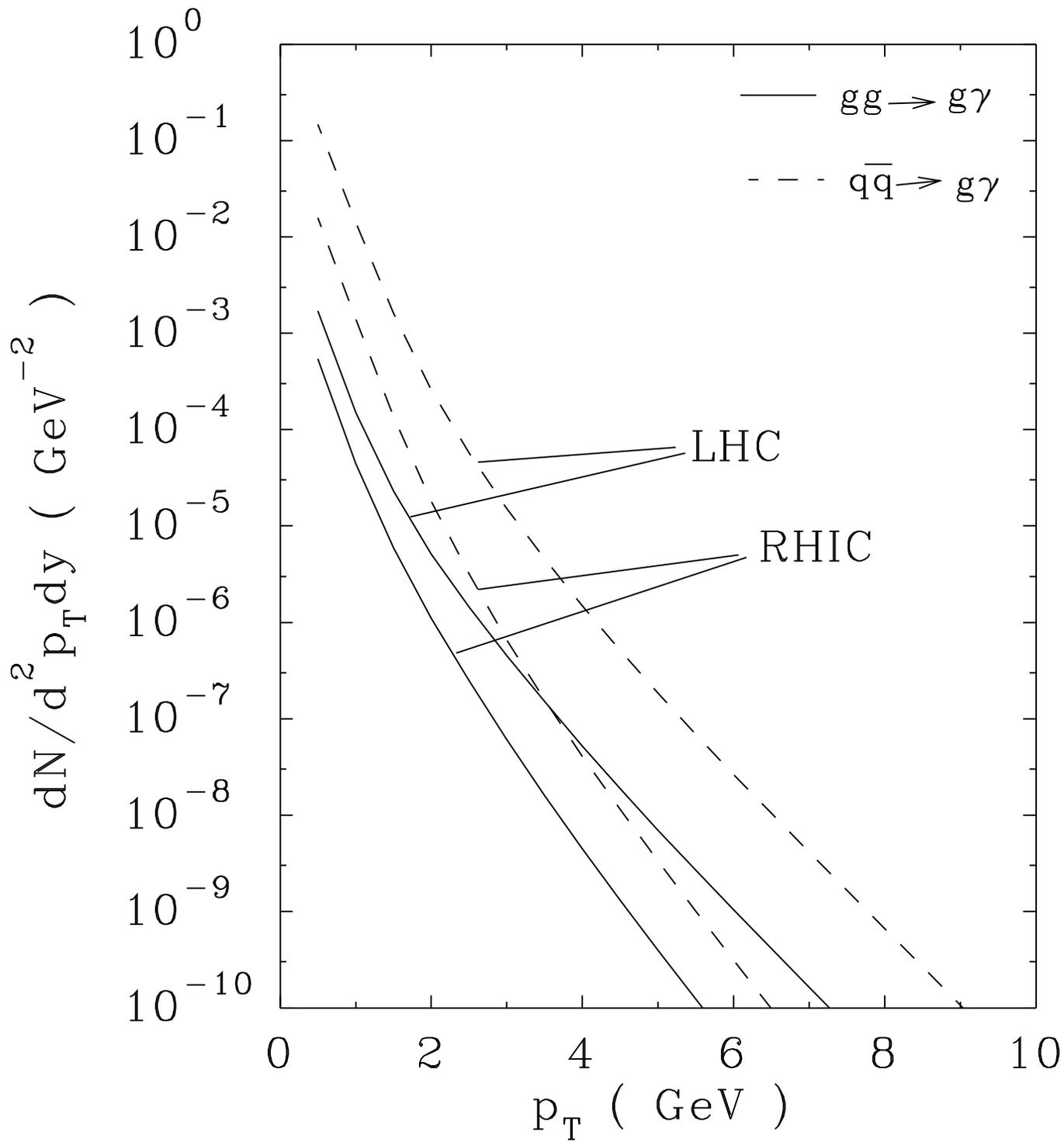
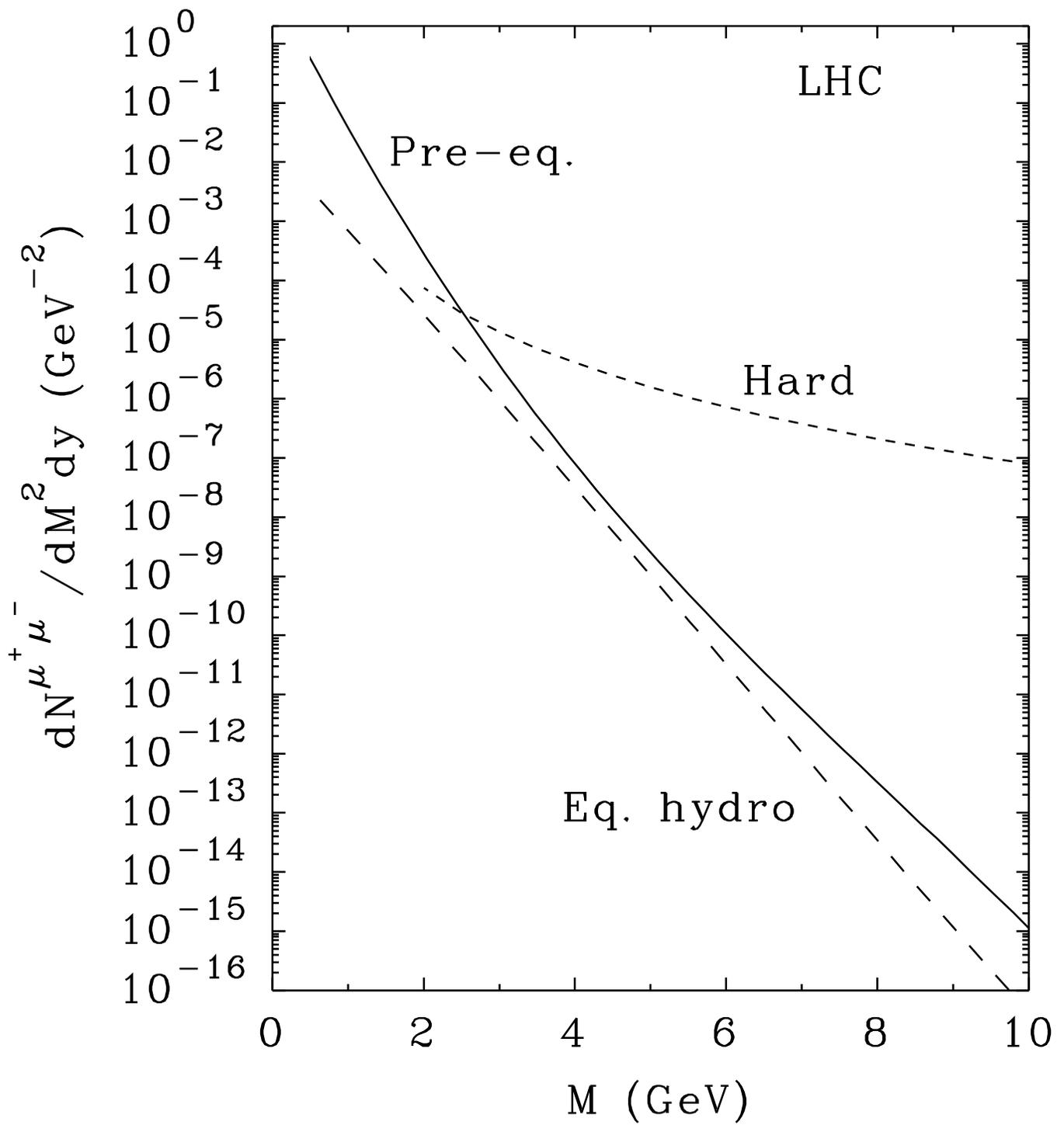
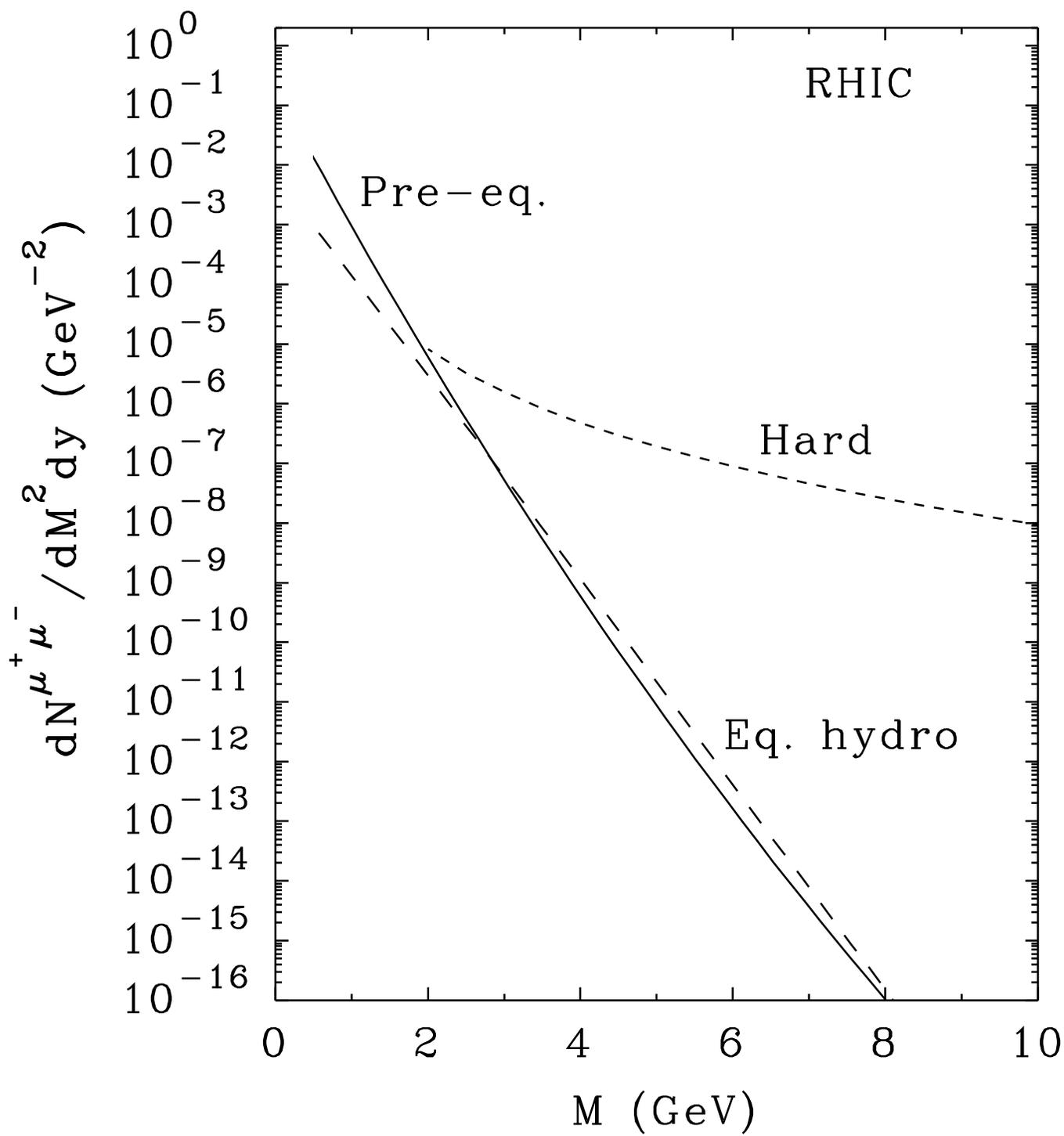
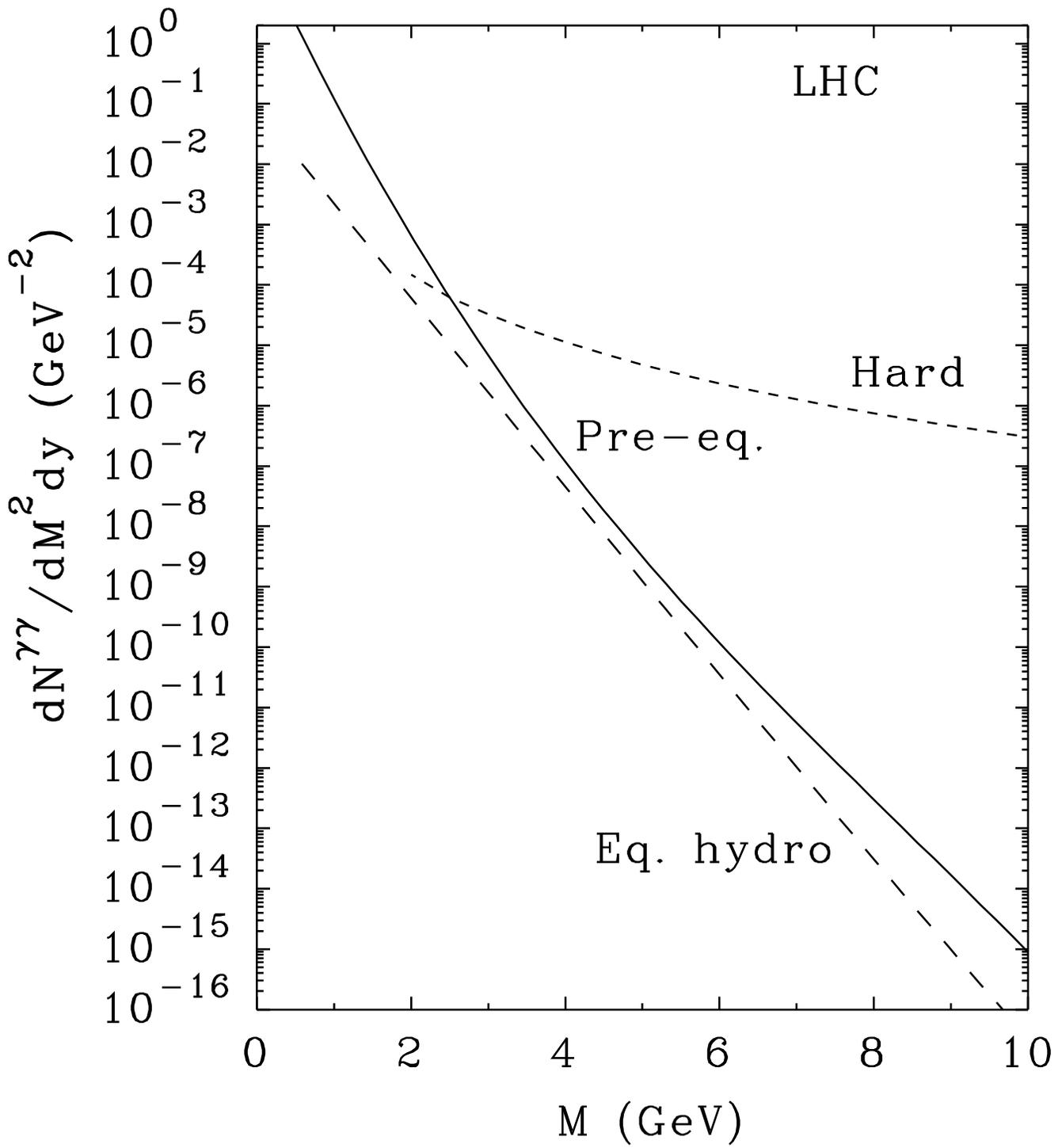
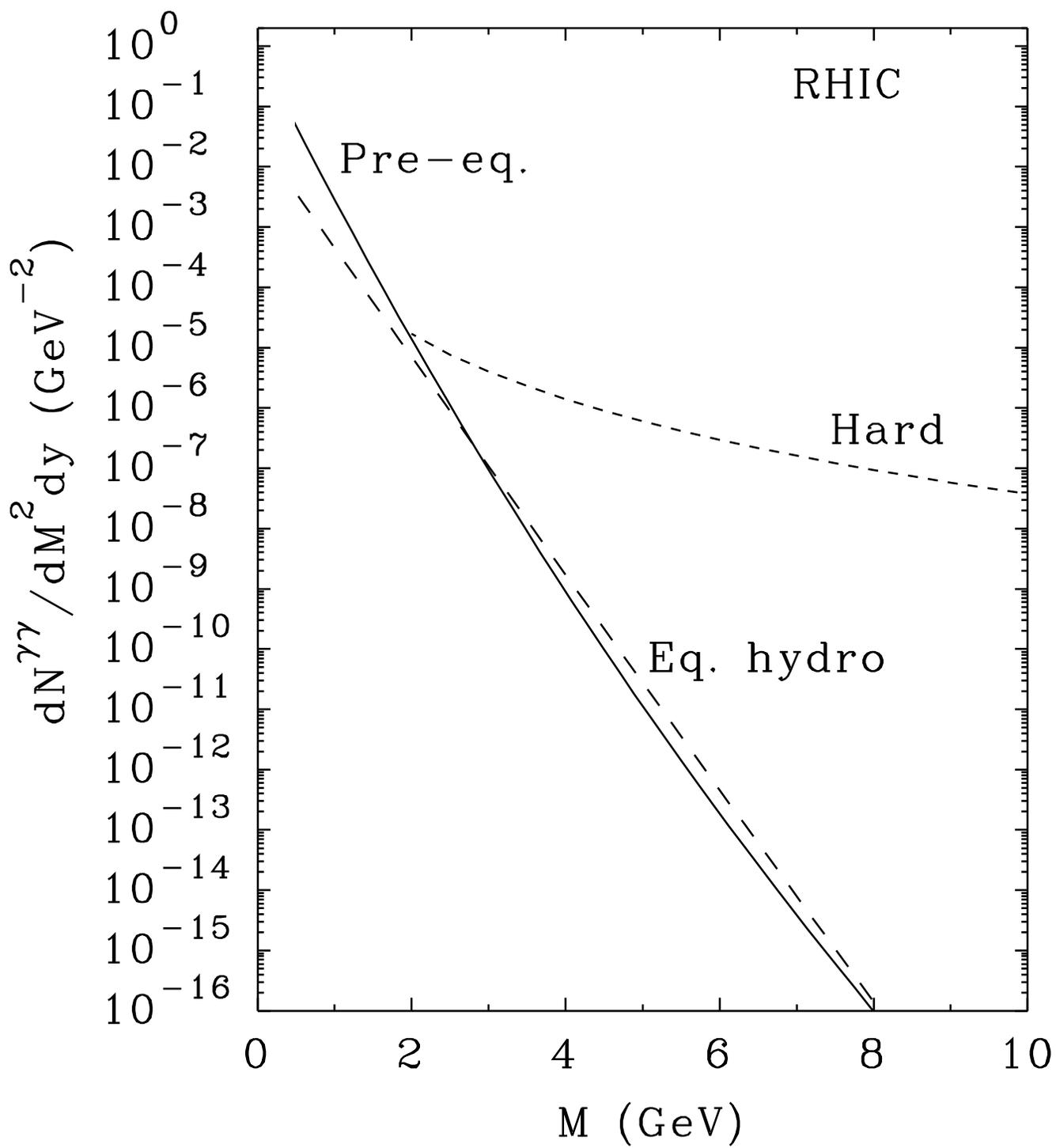
\end{document}